\documentclass[sigconf,natbib=true,anonymous=false]{acmart}

\acmConference[SIGIR ReNeuIR'22]{The ACM SIGIR 2022 Workshop on Reaching Efficiency in Neural Information Retrieval (ReNeuIR)}{July 15, 2022}{Madrid}
  
\usepackage{algorithm}
\usepackage{algpseudocode}
\usepackage{amsmath}
\usepackage{graphics}
\usepackage{epsfig}
\usepackage{multirow}
\usepackage{float}
\usepackage{geometry}
\usepackage{graphicx}
\usepackage{subfigure} 
\usepackage{footmisc}
\usepackage{balance}

\newcommand{\ali}[2]{\makebox[#1][l]{#2}}

\usepackage{color}

\begin{document}
\begin{sloppypar}
\title{SEINE: SEgment-based Indexing for NEural information retrieval}

\author{Sibo Dong}
 \email{sd1242@georgetown.edu}
 \affiliation{%
 \department{InfoSense, Dept. of Computer Science}
  \institution{Georgetown University, USA}
  \country{USA}
  }

\author{Justin Goldstein}
 \email{jjg130@georgetown.edu}
 \affiliation{%
 \department{InfoSense, Dept. of Computer Science}
  \institution{Georgetown University, USA}
  \country{USA}
  }
  
  \author{Grace Hui Yang}
 \email{grace.yang@georgetown.edu}
 \affiliation{%
 \department{InfoSense, Dept. of Computer Science}
  \institution{Georgetown University, USA}
  \country{USA}
  }

\begin{abstract}

Many early neural Information Retrieval (NeurIR) methods are re-rankers that rely on a traditional first-stage retriever due to expensive query time computations.  Recently, representation-based retrievers have gained much attention, which learns query representation and document representation separately, making it possible to  pre-compute document representations offline and reduce the workload at query time. Both dense and sparse representation-based retrievers have been explored. However, these methods focus on finding the representation that best represents a text (aka metric learning) and the actual retrieval function that is responsible for similarity matching between query and document is kept at a minimum by using dot product. 
One drawback is that unlike traditional term-level inverted index,  the index formed by these embeddings cannot be easily re-used by another retrieval method. Another  drawback is that keeping the interaction at minimum hurts retrieval effectiveness. 
On the contrary, interaction-based retrievers are known for their better retrieval effectiveness. In this paper, we propose a novel SEgment-based Neural Indexing method, SEINE, which provides a general indexing framework that
can flexibly support a variety of interaction-based neural retrieval methods.  We emphasize on a careful decomposition of common components in existing neural retrieval methods and propose to use segment-level inverted index to store the atomic query-document interaction values. Experiments on LETOR MQ2007 and MQ2008 datasets show that our indexing method can accelerate multiple neural retrieval methods up to 28-times faster without sacrificing much effectiveness.





\end{abstract}


\maketitle

\section{Introduction}




Neural Information Retrieval (NeurIR) methods use deep neural networks to rank text documents in response to user queries \cite{mitra2018introduction} and  achieve state-of-the-art performance for ad hoc retrieval \citep{huang2013learning, shen2014learning, clinchant2013aggregating, lu2013deep, gupta2014query, ganguly2015word, guo2016deep, grbovic2015search, mitra2015exploring, zheng2015learning,yang2019end, izacard2020leveraging, severyn2015learning,nogueira2019passage,khattab2020colbert}. 
However, most NeurIR methods are re-ranking methods, which must rely on a first-stage  conventional retriever such as BM25~\cite{robertson2009probabilistic} to obtain a manageable set of  relevant document candidates. They re-rank this set because for NeurIR methods to rank the entire corpus is  computationally prohibitive. 

To resolve the efficiency issue, representation-based NeurIR approaches have been explored for they can pre-process document collections offline and offload much of the computation burden. Representation-based methods encode a query or a document into a single fixed-size vector. Each query and each document has its own embedding vector, trained in a way that the embeddings can be used in dot product to measure query-document similarity. 
This setup allows pre-computation of document representations of the entire collection at offline time. Compared to a conventional retrieval system that consists of an indexing phase and a retrieval phase \cite{spider}, representation-based approaches' learning and storing the document embeddings can be thought of the ``indexing" phase; and calculating and sorting the dot products is the ``retrieval" phase. 
It can be shown in the following pipeline (Here $E(q)$ and $E(d_i)$ are embeddings for query $q$ and document $d_i$, respectively): 

\begin{equation}\footnotesize
\label{eq:representation-based}
\begin{aligned}
    \textrm{Learn } E(d_i) & \rightarrow \textrm{Store } E(d_i) \textrm{ in index}, \forall i & \searrow & & \\
    & & & \forall i, E(q) \cdot E(d_i) \rightarrow & \textrm{Top-k results  } \\
    \textrm{Learn } E(q) & \rightarrow E(q) & \nearrow & & \textrm{by MIPS, ANN.}
\end{aligned}
\end{equation}

The embedding vectors can be either dense or sparse. Dense vectors are usually short, with nearly all non-zero entries. Sparse vectors, on the contrary, can be long and with many zero-valued entries. 
Dense, representation-based retrievers (e.g., DPR\cite{dpr}, SBERT~
\cite{sbert}, Condenser~\cite{condenser}, ICT~\cite{ict}, RocketQA~\cite{rocketqa}, ANCE~\cite{ance}, RepBERT\cite{repbert}, and  ColBERT~\cite{khattab2020colbert}) have gained much attention recently. They study pre-training or fine-tuning (mostly fine-tuned from the BERT embeddings~\cite{devlin2018bert}) methods to obtain dense low-dimensional encodings for queries and documents. Unlike results obtained from the sparse bag-of-words (BoWs) representations,  top-K results obtained from these dense representations cannot be efficiently found without any approximation. Instead, they must be assisted with efficient search algorithms for approximate nearest neighbor search (ANN)~\cite{faiss,ance} or maximum inner-product search (MIPS)~\cite{mips}. Popular efficient search algorithms leverage ideas such as  hashing (e.g. LSH~\cite{lsh}), clustering~\cite{10.1109/TPAMI.2018.2889473}, product quantization (e.g. ScaNN \cite{scann}, FAISS \cite{faiss}) and dimension reduction (e.g. PCA \cite{pca}, t-SNE \cite{tsne}), equipped with carefully chosen data structures. 





Sparse, representation-based retrievers have also been explored to further improve indexing and retrieval efficiency by forcing more zero entries in the embeddings. Existing methods to enforcing sparsity include using gating to select entries (e.g., SparTerm \cite{sparterm}), assigning zeros to non-query terms (e.g., EPIC \cite{epic}), and  regularizing the vectors by minimizing flop operations (e.g., SPLADE \cite{splade} and FLOPS \cite{flops}), etc. Although these methods make the embedding vectors sparser, they do not change the basics of the dual-encoder setup (illustrated in Eq.~\ref{eq:representation-based}). Most of them still use the over-simplified dot product 
as their retrieval function (e.g., SparTerm \cite{sparterm}, EPIC \cite{epic}, and  SPLADE \cite{splade}). It is unclear how the benefit gained from sparse representations can be leveraged to employ more efficient data structures such as the inverted index.


We argue that representation-based methods, including both dense and sparse ones, have a few drawbacks for document retrieval. First, unlike traditional term-level inverted index, the index formed by a representation-based approach's embeddings cannot be easily re-used by another retrieval method. Representation-based methods focus so much on finding the representation that best represents a piece of text, aka learning the metric space as in metric learning. Every index that is learned by a representation-based method is unique to itself, which makes it hard to be re-usable by others. When one works on a representation-based method, each time she must process the document collection and rebuild the entire index. It contradicts a common expectation for an index -- that it should be general and re-usable by down-stream tasks such as various retrieval methods.  


Second, in a representation-based method,  the actual retrieval function responsible for similarity matching between query and document is kept to the bare minimum. Most of them use a dot product.  ColBERT~\cite{khattab2020colbert} used a slightly more  sophisticated  maximum cosine similarity function,  but it is still over-simplified for expressing the interaction between query and document. In fact, researchers have discovered that separating query and document in the indexing phrase and keeping their interactions at minimum hurts retrieval effectiveness. Wang et al. pointed out that it is necessary for dense, representation-based retrievers to interpolate with sparse, interaction-based retrievers such as BM25~\cite{10.1145/3471158.3472233} to gain better performance. Luan et al. proved that embedding size imposes an upper bound on effectiveness that a dense retriever can achieve and on text length that it can handle~\cite{10.1162/tacl_a_00369}. It explains why few successes for these approaches have been seen on first-stage, long document retrieval.



On the other hand, interaction-based  retrievers have been known for their superior retrieval effectiveness. This can be witnessed by the long-time  record set by BM25, a sparse, interaction-based method in the pre-neural era, and by the top performance set by MonoBERT \cite{nogueira2019passage}, an all-in-all, dense interaction-based method marked by its extensive interactions among query terms and document terms. Many early NeurIR methods are also sparse, interaction-based methods~\citep{guo2016deep, pang2016text, pang2016study, pang2017deeprank, xiong2017end, fan2018modeling, robertson2009probabilistic, tang2019deeptilebars}. When receiving a query from the user, they generate a query-document interaction matrix in real-time and feed it into neural networks to predict a relevance score. Early interaction-based neural models do not have an index and must construct a large interaction matrix at query time, which is the main obstacle that prohibits them from succeeding in applying to first-stage, full-length documents. 

A few pieces of work have been proposed to support indexing for  interaction-based neural retrievers. Most of them take advantage of an inverted index for its fast lookup functions. They share a  pipeline illustrated as the following:
\begin{equation} \footnotesize
\label{eq:interaction-based}
\begin{aligned}
    \textrm {Process } d_i & \rightarrow \textrm{Store} \langle v, d_i\rangle \textrm{ in index}, \forall i & \searrow & & \\
    & & &  \textrm{Lookup } q \textrm{ in } v \rightarrow & \textrm{Top-k results}\\
   \textrm {Process } q  &\rightarrow q & \nearrow & & \textrm{by } s(q,d_i),
\end{aligned}
\end{equation}
where $v$ is the vocabulary obtained from the document collection, $\langle v, d_i\rangle$ is the interaction between $v$ and document $d_i$, and $s(q,d_i)$ is a retrieval function that measures the similarity between $q$ and $d_i$. 

For instance, DeepCT~\cite{deepct} substituted term frequency with context-aware term weights aggregated from BERT embeddings~\cite{devlin2018bert}, and stored them in an inverted index. Their followup work, HDCT \cite{hdct}, succeeded in full-length document retrieval. TILDE~\cite{tilde} stored conditional term probabilities in an inverted index to support deep query-likelihood calculations in its retrieval function. SPARTA~\cite{sparta} stored dot products between vocabulary terms to document terms in an inverted index.  


While it is encouraging to see these methods build indices for sparse, interaction-based retrievers, we find their indices are tailored to the specific neural retrieval function that they use in their retrieval phase. 
It is suboptimal if these indices cannot be general enough to be re-used by other neural retrievers. For instance, neural retrievers developed prior to BERT, including KRNM~\cite{robertson2009probabilistic}, HiNT~\cite{fan2018modeling}, and DeepTileBars~\cite{tang2019deeptilebars}, are  effective on document retrieval but have no index to support them. SNRM can be made to support them, as we demonstrated in our experiments, however, the matchings are done with latent terms, not actual lexical terms and the effectiveness degrades much.

In this paper, we propose a novel {\bf SEgment-based Neural Indexing method}, \textbf{SEINE}, which provides a general indexing framework that
can flexibly support a variety of  neural retrieval methods. We focus on facilitating interaction-based retrieval methods for their higher effectiveness. During query time, a retriever can only look up pre-computed interaction function values for corresponding query terms from the index, quickly merging them into a query-document interaction matrix and sending it to neural networks to predict a relevance score. Moreover,  we adopt a flexible, segment-based design in our index to support  query-document interaction at different granularities. For instance, the query-document interaction can be done at the document-level (e.g. BM25 \cite{robertson2009probabilistic} and TILDE~\cite{tilde}), term-level (e.g., KRNM~\cite{robertson2009probabilistic}), or topical segment-level (e.g., HiNT~\cite{fan2018modeling} and DeepTileBars~\cite{tang2019deeptilebars}). We believe as long as we can decompose the retrieval methods and identify the interaction units used between the query and the document, we should be able to build an index that is general, modularized, and reusable.  
Our segment-level inverted index stores common components, which we call atomic interaction values, including term frequency, BERT embedding, conditional probabilities, etc.  We also leverage Spark programming to accelerate the entire indexing process. Experiments on LETOR MQ2007 and MQ2008 datasets show that our indexing method can accelerate multiple neural retrieval methods up to 28-times faster without sacrificing much effectiveness.

\begin{figure*}[t!]
  \centering
  \includegraphics[width=0.8\linewidth]{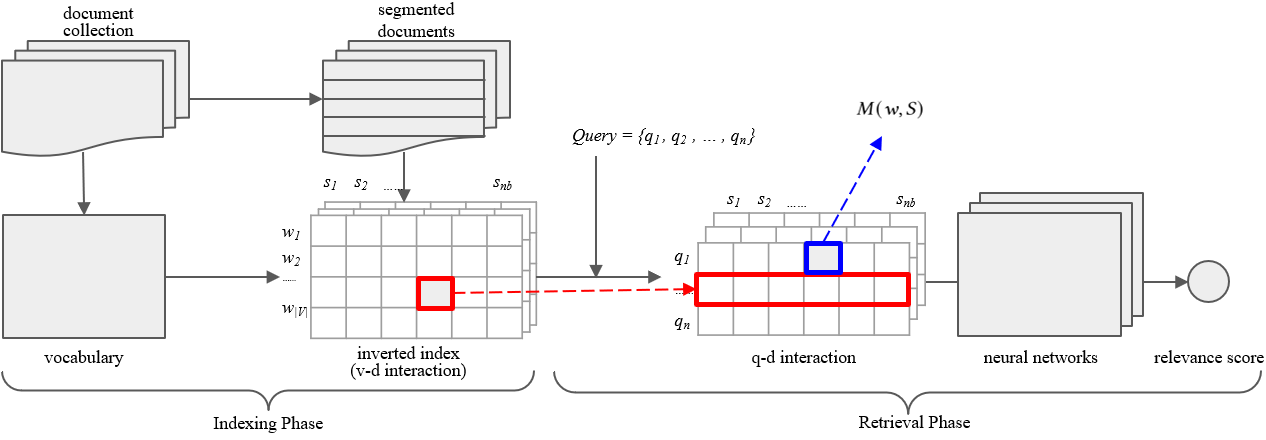} \vspace{-0.07in} 
  \caption{SEINE: SEgment-based Indexing for NEural information retrieval.}
  \label{fig:framework} 
\end{figure*}


Note that the most effective neural retrievers at the moment when the paper is being written are dense, interaction-based methods, such as MonoBERT~\cite{nogueira2019passage} and monoT5~\cite{monot5},  whose all-in-all interactions within its transformer blocks cannot be easily decomposed. We leave creating indices for this type of retrievers as future work and focus on sparse, interaction-based methods.



Our work makes the following contributions: 
\begin{itemize}
    \item We propose a general indexing framework for constructing  reusable indices to support sparse, interaction-based neural retrievers;  
    \item We design our index to enable multiple sparse, interaction-based NeurIR methods that previously have no index and suffer from high latency; Our work rejuvenates them for first-stage, full-length document retrieval; 
    \item We also demonstrate how to make use of the Spark programming to accelerate the entire indexing process.
\end{itemize}

\section{Method}













Our focus is on decomposing existing interaction-based neural retrievers by two measures: 1) recognizing that they perform query-document interactions at different granularities and providing a flexible segment-based approach to suit them all; 2) identifying a list of atomic interaction function values     that can be pre-computed and stored in an inverted list to support common neural retrievers. 

Figure \ref{fig:framework} illustrates SEINE's framework. It supports end-to-end ad hoc document retrieval and has two phases, an indexing phase and an retrieval phase. The \textbf{indexing phase} is query-independent and can be done offline. During indexing,  we (1) process the entire corpus and obtain a vocabulary $v$, (2) segment each document in the collection, and then  (3) create an inverted index by interacting each vocabulary term $w_i$ with each document $d_j$ on the segment level. We call this interaction {\it v-d interaction} 
meaning each vocabulary word interacting with each document in the collection. The inverted index is used to store a list of interaction values for each word-segment pair. The number of segments per document is standardized for all documents. (4) We also adopt   Spark\footnote{\url{https://spark.apache.org/}.} and make use of its parallel programming to accelerate the indexing process. At the {\bf retrieval phase}, when a query $q$ comes in, each query word is looked up from the v-d interaction matrix stored in the index and the matched terms'  rows are extracted and stacked into another interaction matrix called {\it q-d interaction}. 
We feed the q-d interaction matrix into a neural retrieval function to obtain the final relevance scores $s(q,d_j), \forall j$. Since the number of query terms are small, calculations at the retrieval phase is sparse and quick to finish. 




\subsection{Pre-Processing a Document Collection} 


 
To begin, we process the entire corpus $C=\{d_1, d_2, ..., d_{|C|}\}$ and obtain a vocabulary $v=\{w_1, w_2, ... , w_{|v|}\}$. 
The vocabulary is the set of unique terms appearing in the corpus.  Standard text pre-processing steps are taken to attain the vocabulary. First, we tokenize all documents in the collection $C$ into a sequence of tokens using the Wordpiece tokenizer~\cite{wu2016google}. Then, we remove the most frequent 10\% and the least frequent 10\%  terms from the vocabulary based on their collection-level frequency. It is to exclude misspellings, rare words, programming scripts, punctuation, special symbols, stopwords, etc.  While we take this pass on the entire document collection, we keep track of the inverse document frequency of each term ($idf(w_i), w_i \in v$) and store them: $
idf(w_i) = \log \frac{|C|}{|\{j|w_i \in d_j\}|+1}$.


For most neural retrievers that we support, we assume that all incoming query terms can be found in the vocabulary, so that at query time the {\it q-d interaction} matrix can be built by looking up in the pre-computed {\it v-d interaction} matrix, instead of calculating the {\it q-d interactions} on the fly. We are aware that this assumption may not be valid and out-of-vocabulary terms can certainly be in a query. 
In some neural indexers that we compare with (e.g. SNRM \cite{zamani2018neural}), however, a latent semantic representation is used as the indexing unit, which can be seen as a way to alleviate the vocabulary mismatch problem.

\subsection{Segmenting Documents to Support Various Interaction Granularities} 
Existing interaction-based neural retrievers perform query-document interaction at different granularities. Some are performed at the document-level (e.g. SNRM  ~\cite{zamani2018neural} and BM25 \cite{robertson2009probabilistic}), which interacts a query $q$ with an entire document $d$; some are at the term-level (e.g., KRNM~\cite{robertson2009probabilistic}), meaning interacting a single query term and a single document term; and others are at segment-level  that represent topics in a document (e.g., HiNT~\cite{fan2018modeling} and DeepTileBars~\cite{tang2019deeptilebars}). We propose  a flexible segment-based indexing approach to support query-document interaction at different granularities. Our chosen interaction unit is the segment, whose size can be adjusted to include term- and document-level interactions.

For a document $d$, we split it into non-overlapping segments following the TextTiling algorithm~\cite{hearst1994multi}. The segmentation is done based on the similarity between fixed-sized, neighbouring text windows. If any two neighboring windows show high similarity, we merge them into a bigger segment. If they are dissimilar, they are put into different segments. 
The resulting segments roughly reflect topical boundaries in a document. Unsurprisingly, the number of segments $y$ in one document can be quite different from that in another. To standardize $y$ for different documents, we (1) pad empty segments if  $y<=n_b$, a pre-defined, adjustable dimension parameter, or (2) squeeze all remaining text into the final segment if $y>n_b$. Eventually, all documents contain an equal number of $n_b$ segments: $d = \{S_1, S_2, ..., S_{n_b}\}$.  When $n_b = 1$, it is equivalent to interacting at the document-level; when $n_b = |v|$, it is equivalent to interacting at the term-level. Note that driven by a completely different motivation, \cite{10.1162/tacl_a_00369} also spoke about the advantage of using multiple, fixed length vectors, instead of a single vector, to represent a document. Here we introduce segment-level indexing for its flexibility to cover query-document interactions at all levels of granularities, when $n_b$ is set to different values. 

\subsection{Storing Atomic Interactions in Inverted Index} \label{matrix}



In this paper, we propose to decompose components in existing neural retrievers  that are related to query-document interactions into atomic interaction functions  and store their values into the index. These atomic interactions include term frequency, operations over BERT embeddings, kernel functions, conditional probabilities, etc. 

For each term $w \in v$ and an interaction unit $S$, where $S$ is a text segment (which can be adjusted to represent a document or a term), we pre-calculate and store the following atomic interaction function values:  
\begin{itemize}
    \item{\bf Term frequency: } $tf(w, S)$, the number of occurrences of term $w$ in $S$. It can be stored to support traditional retrieval methods such as BM25~\cite{robertson2009probabilistic} and neural retrievers such as DeepTileBars~\cite{tang2019deeptilebars}. 
    
    \item {\bf Indicative inverted document frequency:} $idf(w)\times\mathbb I_S(w)$,  where  $idf(w)$ is the inverse document frequency of term $w$ and $\mathbb I_S(w)$ is indicates whether $w$ is in $S$. This function can be used to support traditional retrieval methods such as BM25~\cite{robertson2009probabilistic} and the neural retrievers such as HiNT~\cite{fan2018modeling} and DeepTileBars~\cite{tang2019deeptilebars}. 
    
    \item {\bf Dot product:} $ \sum_{t \in S} E(w) \cdot E(t)$, where $E(.)$ is an embedding output from a pre-trained neural encoder, such as word2vec  \cite{mikolov2013distributed} or BERT \cite{devlin2018bert}. The dot product measures similarity between the two embeddings.  In theory, $E(.)$ can be any dense or sparse representations for a text sequence. Therefore, this interaction function can be used to store interaction between the dense representations of a word and a segment. It is thus can be used to support  MatchPyramid~\cite{pang2016study} and dense retrievers such as COIL \cite{coil}. 
    
    \item{\bf Cosine similarity: } $\sum_{t \in S} \frac{E(w) \cdot E(t)}{|E(w)| \cdot |E(t)|}$, similarly to the dot product and used as another similarity function. This function supports neural retrievers such as KNRM~\cite{xiong2017end} and HiNT~\cite{fan2018modeling}.
    
    \item{\bf Gaussian kernel: } $\max_{t \in S} exp(-(E(w)-E(t))^2)$, proposed by \cite{xiong2017end} to measure the distance between two terms within a semantic neighborhood. It can be used to find the most similar  synonym to a word in $S$. It supports KRNM~\cite{xiong2017end} and DeepTileBars~\cite{tang2019deeptilebars}. 
    

 \item{\bf Linear aggregation on BERT Embeddings: } $a\cdot E_w(S) + b$, where $E_w(S)$ is a vocabulary term $w$'s BERT embedding in text $S$. It  linearly combines BERT embeddings using learned weights $a, b$ and can be thought of an aggregated contextual term weight for $w$ in $S$. It supports DeepCT \cite{deepct} and can be used in combination with traditional retrievers such as BM25. 
  
  \item{\bf Max operation on BERT Embeddings: } $\max_{t\in S} f_S(E(t))\cdot E(w)$, where $f_S$ is the logarithm of the softplus over BERT embeddings. $E(\cdot)$ is a BERT embedding. This function selects the most similar term in a piece of text to a vocabulary term ($w$) and records their similarity. It can be used to support EPIC~\cite{epic} and ColBERT~\cite{khattab2020colbert}.

 \item{\bf Multi-layer Perceptron on BERT Embeddings: } $MLP(E_w(S))$, where $MLP(\cdot)$ is multilayer perceptron with activations over BERT embeddings. It can be used to support retrievers such as DeepImpact \cite{deepimpact}. 
         
    \item{\bf Log conditional probability: } 
    $\log P_\theta(w|S)$, where $P_\theta(w|S)$ is the conditional probability for term $w$ in text $S$ and can be obtained by using a language modeling head on the [CLS] token of $E_w(S)$, which is a vocabulary term $w$'s BERT embedding in text $S$. It can be used to support deep query likelihood models such as TILDE \cite{tilde}. 
\end{itemize}

Some of these atomic interaction functions are proven to be essential for ad-hoc retrieval and others are widely used in interaction-based neural retrievers. Note that the list of functions is not exhaustive and one can certainly expand the list or choose a subset of functions in their index. One condition must be satisfied to identify those atomic functions -- that the vocabulary entries, in our case the terms, must be  independent of each other. In all-in-all interaction-based methods, such as MonoBERT,  however, terms interact within and across a query and document and the  interactions in the transformer blocks cannot be easily decomposed. We thus do not support them.

We define a vocab-segment interaction column vector $M(w_i,S_k)$ for the $i^{th}$ vocabulary word $w_i$ and the $k^{th}$ segment (across all documents), and keep each of the above atomic interaction function values in it. 
These term-segment interactions $M(w_i,S_k)$ are then  form the v-d interaction matrix: 
\begin{equation}  \small 
M_{v, d}=concat\{\ M(v,S_1),\ M(v,S_2),\ ...\ ,\ M(v,S_k),\ ...\ ,\  M(v,S_{n_b})\ \} \end{equation} where $M(v,S_k)$ is obtained by combining for all terms $w_i \in v, \forall i$ in the $k^{th}$ segment. 
Eventually, we generate an interaction matrix of dimension $|v| \times n_b \times n_f $ which stores the atomic interactions for every vocab term-document pair, where $|v|$ is the vocabulary size, $n_b$ is the number of segments in a document, and $n_f$ is the number of atomic interaction functions.  

At retrieval time, we obtain the q-d interaction matrix for an incoming query by looking up query terms in the vocabulary and stacking the rows in the index that stores their pre-computed interaction scores:  
\begin{equation} \small 
M_{q,d} = stack_{w_i \in q \cap v}\{\ M_{w_1,d},\ M_{w_2,d},\ ...\ ,\ M_{w_i,d}\ ... \}. 
\end{equation}


Our index can support neural interactions at different granularities, only by varying the segment size. To support neural retriever with document-level interactions, we 
let the segment size equal the document length so that there is one segment and it is $d$. The v-d interaction matrix becomes $M_{v, d} = M(v, d)$.
To support neural retrieval methods with  term-level interactions, we treat each term as a segment, i.e. the segment size is one term. The v-d interaction matrix becomes: $M_{w_i, d} = concat\{\ M(w_i,t_1),\ M(w_2,t_2),\ ...\ ,\ M(w_i,t_{n_d})\ \}$, and $n_d$ is the document length.



\subsection{Accelerating with Spark} 


\begin{algorithm} [t]
\caption{Spark pseudo-code for indexing.}
\label{alg:spark} \small
\begin{algorithmic}[1] 
\State Initialize Spark environment and configuration
\State Import functions segmentation, interaction
\State \ali{3em}{Vocab} $\leftarrow RDD\{w_1, w_2, ... , w_{|V|}\}$  \Comment{create RDD}
\State \ali{3em}{Corpus} $\leftarrow RDD\{d_1, d_2, ... , d_{|C|}\}$  \Comment{create RDD}
\State \ali{3em}{Segmts}   $\leftarrow$ Corpus.{\it map }(segmentation)   \Comment{document segmentation} 
\State \ali{3em}{Cart} $\leftarrow \textrm{Vocab.}cartesian(\textrm{Segmts})$
\State \ali{3em}{Index} $\leftarrow$ Cart.{\it map }(interaction)    \Comment{calculate $M$ as in \S~\ref{matrix}}
\State \ali{3em}{Index} $\leftarrow$ Index.{\it filter }($tf>\sigma_{index}$)
\State \ali{3em}{Index} $\leftarrow$ Index.{\it reshape }          \Comment{v-S to v-d }
\State Index.{\it saveAsPickleFile }()
\end{algorithmic}  
\end{algorithm}

Large corpora can contain tens of millions of documents and a vocabulary of tens of thousands of words. Given a dozen to three dozens of segments in a document, there can be trillions of query-segment interactions when building the index. In our implementation, we leverage Spark~\cite{zaharia2010spark} to accelerate the indexing process. Spark uses a special data structure called the resilient distributed dataset (RDD), which can hold a large  distributed collection of objects and has a built-in parallel programming infrastructure. Each RDD automatically splits into multiple data partitions and can be computed on different computer nodes~\cite{karau2015learning}. Spark has two types of programming functions, transformations and actions, and uses a lazy evaluation mechanism. Transformations are not real computations but prototyping functions that wait for computation paths optimization and data parallelization are done by the underlying Spark infrastructure. Actions do the actual computations, but only when it is absolutely necessary to compute. 

We employ Spark to accelerate the indexing process. After obtaining the vocabulary $v$ and segmenting all documents into segments, we perform v-d interaction over all the terms in $v$ and all segmented documents in corpus $C$: 
(1) We create RDDs for both vocabulary term list and document list. 
(2) We use a transformation operation cartesian($\cdot,\cdot$) to compute a Cartesian product between two RDDs, for example we have a vocabulary $V=\{apple, banana\}$ and a collection $C=\{d1, d2\}$, the Cartesian function returns a list of term-document pairs: $\{(apple, d1), (apple, d2), (banana, d1), (banana, d2)\}$. 
(3) We use another transformation operation map($\cdot$) to calculate interaction matrix by applying function interaction($\cdot$) to v-d pairs, where interaction($\cdot$) is defined in Section 2.3. 
(4) In order to balance the memory storage with information loss, we use a transformation operation filter to control the index sparsity by the threshold $\sigma_{index}$. For example, if $\sigma_{index}=0$, only the documents containing the corresponding term are stored in the index. 
With $\sigma_{index}>0$ can further improve the sparsity and efficiency of the index, but it may lead to information loss and effectiveness drop. 
(5) We combine and reshape the word-segment interaction into a v-d interaction matrix.
(6) We then use an action operation to write the results into disk. Algorithm 1 outlines our Spark implementation. 

\section{Experiments}



\begin{table*}[htb]
\small
\caption{Retrieval Effectiveness and Efficiency on MQ2007 and MQ2008. 
$^*$ denotes statistically significant degradation on effectiveness and $^\dagger$ for statistically significant improvement on efficiency compared to corresponding retrieval method with ``No Index".}  
\label{tab:result} 
\subtable[\textbf{MQ2007.}]{
\begin{tabular*}{\textwidth}{@{\extracolsep{\fill}}*{2}{l}*{5}{l}*{4}{r}}
\toprule
    \multirow{3}{*}{\bf Indexing} &  \multirow{3}{*}{\bf Retrieval}  &  \multicolumn{5}{c}{\bf Effectiveness}  &  \multicolumn{4}{c}{\bf Efficiency}\\ 
\cline{3-7} \cline{8-11} 
    &  & \multicolumn{1}{c}{P@5}  & \multicolumn{1}{c}{P@10}  & \multicolumn{1}{c}{MAP} & \multicolumn{1}{c}{nDCG@5} & \multicolumn{1}{c}{nDCG@10}  & \multicolumn{2}{c}{Training (ms)} & \multicolumn{2}{c}{Test (ms)} \\
\cline{2-7} \cline{8-9} \cline{10-11}       
No Index& Dot Product &0.328&0.344&0.418&0.282&0.332&&&& \\
        & KNRM  &0.355&0.382&0.461&0.379&0.412&42.63&&16.70 \\ 
        & HiNT$\diamond$  &0.461&0.418&0.505&0.463&0.490&1139.34&&1009.11 \\
        & DeepTileBars  &0.429&0.408&0.474&0.398&0.434&163.91&&73.68 \\
\cline{2-11}
InvIdx$^\ddagger$& BM25$\diamond$    &0.388&0.366&0.456&0.384&0.414 \\
\cline{2-11}
SNRM    & Dot Product   &0.288$^*$&0.307$^*$&0.368$^*$&0.254$^*$&0.302$^*$ \\
        & KNRM  &0.322$^*$&0.347$^*$&0.417$^*$&0.337$^*$&0.404 &35.56&1.2$\times$&13.17&1.3$\times$\\
        & HiNT  &0.401$^*$&0.358$^*$&0.423$^*$&0.379$^*$&0.402$^*$ &958.03&1.2$\times$&860.76&1.1$\times$ \\
        & DeepTileBars  &0.281$^*$&0.304$^*$&0.359$^*$&0.238$^*$&0.290$^*$ & 131.43 & 1.2$\times$ & 56.97& 1.3$\times$ \\
\cline{2-11}   
SEINE    & Dot Product &0.328&0.344&0.418&0.282&0.332 \\
        & BM25 w/ DeepCT weight &0.315&0.327&0.397&0.266&0.314\\
      & KNRM  &0.342&0.372&0.447&0.374&0.401 &11.67$^\dagger$&3.7$\times$&1.22$^\dagger$&13.7$\times$ \\
        & HiNT  &0.453&0.409&0.492&0.452&0.483 &834.64&1.4$\times$&706.42&1.4$\times$ \\ 
        & DeepTileBars  &0.412&0.404&0.468&0.391&0.427 &22.62$^\dagger$&7.4$\times$&2.67$^\dagger$&28.1$\times$ \\
  \bottomrule
\end{tabular*}
} 
\subtable[\textbf{MQ2008.}]{ 
\begin{tabular*}{\textwidth}{@{\extracolsep{\fill}}*{2}{l}*{5}{l}*{4}{r}}
\toprule
    \multirow{3}{*}{\bf Indexing} &  \multirow{3}{*}{\bf Retrieval}  &  \multicolumn{5}{c}{\bf Effectiveness}  &  \multicolumn{4}{c}{\bf Efficiency}\\ 
\cline{3-7} \cline{8-11} 
    &  & \multicolumn{1}{c}{P@5}  & \multicolumn{1}{c}{P@10}  & \multicolumn{1}{c}{MAP} & \multicolumn{1}{c}{nDCG@5} & \multicolumn{1}{c}{nDCG@10}  & \multicolumn{2}{c}{Training (ms)} & \multicolumn{2}{c}{Test (ms)} \\
\cline{2-7} \cline{8-9} \cline{10-11}  
No Index& Dot Product   &0.333&0.281&0.462&0.411&0.183 \\
        & KNRM  &0.355&0.355&0.472&0.499&0.225 &41.27&&16.86&\\
        & HiNT$\diamond$  &0.367&0.255&0.505&0.501&0.244 &1139.34&&1009.11&\\
        & DeepTileBars  &0.425&0.321&0.567&0.548&0.259 &167.20&&74.96& \\
\cline{2-11}
InvIdx$^\ddagger$  & BM25$\diamond$  &0.337&0.245&0.465&0.461&0.220 \\
\cline{2-11}
SNRM    & Dot Product  &0.380$^\dagger$&0.300$^\dagger$&0.513$^\dagger$&0.483$^\dagger$&0.223$^\dagger$ \\
        & KNRM  &0.303$^*$&0.229$^*$&0.417$^*$&0.417$^*$&0.199$^*$ &36.78&1.1$\times$&12.68&1.3$\times$ \\
        & HiNT  &0.291$^*$&0.209$^*$&0.401$^*$&0.445$^*$&0.221$^*$ &941.09&1.2$\times$&904.16&1.1$\times$\\
        & DeepTileBars  &0.356$^*$&0.291$^*$&0.481$^*$&0.434$^*$&0.200$^*$ &134.10&1.2$\times$&58.14&1.3$\times$ \\
\cline{2-11}
SEINE   & Dot Product   &0.333&0.281&0.462&0.411&0.183 \\
        & BM25 w/ DeepCT weight  &0.307&0.239&0.448&0.400&0.197 \\
       & KNRM  &0.346&0.252&0.462&0.485&0.218 &11.58$^\dagger$&3.6$\times$&1.24$^\dagger$&13.6$\times$ \\
        & HiNT  &0.362&0.250&0.485&0.489&0.236 &828.84&1.4$\times$&687.85&1.5$\times$ \\ 
        & DeepTileBars &0.422&0.322&0.569&0.544&0.255 &22.62$^\dagger$&7.4$\times$&2.67$^\dagger$&28.1$\times$ \\
 \bottomrule
\end{tabular*} 
}
\leftline{\textit{$^\diamond$ Results reported in Fan et al. \cite{fan2018modeling}. $^\ddagger$ Indexing method is not applicable to KNRM, HiNT, and DeepTileBars.} } 
\end{table*} 

We experiment on LETOR 4.0, a widely used benchmark dataset for document retrieval. It uses the Gov2 web page collection ($\sim$2M pages) and two query sets from TREC Million Query (MQ) 2007 and 2008 Tracks \cite{qin2013introducing}.
MQ2007 contains about 1,700 queries and 65,323 annotated documents; MQ2008 800 queries and 15,211 annotated documents. We use the official effectiveness metrics in LETOR. They include Precision (P), Normalized Discounted Cumulative Gain (nDCG) \cite{jarvelin2002cumulated} at various positions, and Mean Average Precision (MAP) \cite{zhu2004recall}.  
For efficiency measures, we report the wall clock time in milliseconds for processing a query-document pair during training and during testing, respectively. The training time calculates the average time used for each sample $(q, d_1, d_2, label)$ per epoch. We compare the time spent by a retriever when there is no index supporting it versus there is an index. For experimental runs with No Index, the training time includes the time spent on generating the interaction matrix; for experimental runs  with an an index, it contains the time to lookup from the index. The test time calculates the average time spent on predicting a score for each  query-document pair $(q, d)$. For all runs, we adopt a five-fold cross-validation and report its results. 

\subsection{Baseline Runs}

We organize the experiments by separating a retriever's indexing method from its retrieval method and report the effectiveness and efficiency of the combinations. All runs in our experiments perform first-stage document retrieval, not re-ranking nor passage retrieval. Note that there are many recent neural retrievers being proposed, we select a few representative ones for their advanced retrieval functions, especially for those early neural methods that had no index nor dense representation to remedy the efficiency issue. Previously, they were limited to act as a re-ranker on passage retrieval tasks. In this work, one of our main purpose is to rejuvenate them to apply to first-stage, full-length document retrieval. We therefore select the following 
\emph{retrieval methods} for our experimentation: 

\noindent {\bf \textbullet \space \textbf{Dot Product}} In ~\cite{zamani2018neural}, they calculated the relevance score by $s(q,d) = \sum_{|q_i|>0} q_id_i$, where $q_i$ are the non-zero elements of $q$ and $d_i$ is the document with non-zero elements in the i-th index.
For our runs with SEINE index, we enable  dot product  and score the relevance by summing over all query terms.

\noindent {\bf \textbullet \space \textbf{BM25\cite{robertson2009probabilistic}}:} We compare BM25 results using an inverted index with conventional bag-of-words term weights vs. using the inverted index with one of SEINE's interaction values turned on. The interaction  is the BERT-based term weight proposed by DeepCT \cite{deepct}. 

\noindent {\bf \textbullet \space KNRM~\cite{xiong2017end}:}
an interaction-based neural retrieval method that performs  term-level interaction over term embeddings, then uses  kernel-pooling to extract multi-level soft matching features to learn the relevance score.

\noindent {\bf \textbullet \space HiNT~ \cite{fan2018modeling}:} a hierarchical neural retrieval method that generates segment-level interaction matrices as the network's input, then uses a local matching layer and a global decision layer to learn the relevance score. 

\noindent {\bf \textbullet \space DeepTileBars~\cite{tang2019deeptilebars}:} 
a neural retrieval method that segments documents by topics then scans them by multiple varied-sized Convolutional neural networks (CNNs) to learn the relevance score.

\vspace{0.07in} 

\noindent For the \emph{indexing methods}, we experiment on:  \\

\noindent {\bf \textbullet \space No Index:} This is the case when a retriever directly processes the document collection and interacts a query to all documents at query time. Most neural retrievers are of this type, including the most effective MonoBERT~\cite{nogueira2019passage}. In our experiments, we test on the above-mentioned neural retrievers to illustrate.  

\noindent {\bf \textbullet \space Inverted Index (InvIdx):} the traditional indexing method using the bag-of-words representation. It stores a posting list for each vocabulary term $w$, consisting all the $(d, tf(w,d))$ pairs for every document $d$ containing $w$. Note that InvIdx does not store any semantic interactions nor embedding-based interactions, so it cannot support the recent neural retrieval methods. 
    

\noindent {\bf \textbullet \space SNRM~\cite{zamani2018neural}:} a neural indexing method that learns a sparse latent representation for the interaction between each pair of training query and document, and stores the latent nodes in an inverted index. Note that in SNRM, the latent words are independent of each other, which makes it satisfy the condition that we mentioned in Section 2.3 that vocabulary entries must be independent of each other. We can therefore use the latent nodes as vocabulary entries in the inverted index and manage to apply SNRM to KNRM, HiNT, and DeepTileBars. We first represent the query and document as a sequence of latent words, and then generate the interaction matrix based on the latent words.  The average document length is significantly reduced  from 1500 words to 130 latent words on the ClueWeb corpus.

\noindent {\bf \textbullet \space SEINE:} The segment-based neural indexing method proposed in this paper. Note that for different retrieval methods, SEINE turns on a different subset of atomic interaction functions. For instance, DeepTileBars are supported by term frequency, indicative inverted document frequency and Gaussian kernel values; KNRM are supported by cosine similarity. 

\vspace{0.07in} 

We use the following parameter settings in our implementations. To keep a manageable vocabulary size, we use the middle 80\% frequent terms in the vocabulary. For LETOR 4.0, the vocabulary is around 40,000 words. We set $\sigma_{index}=0$, which means documents containing the corresponding term and only them are stored into the index. With $\sigma_{index}>0$ can further improve the sparsity and efficiency of the index, but it may lead to information loss and effectiveness drop; so we set it to be zero. For all experimental runs, whenever possible, we use the default settings recommended in their original published papers. We employ the pre-trained word2vec model \cite{mikolov2013distributed} to represent terms  for KNRM, HiNT, and DeepTileBars, and BERT embeddings in BERT-base-uncased for DeepCT. 


\begin{figure}[t]
    \centering
    \subfigure[]{
    \includegraphics[width=0.47\linewidth]{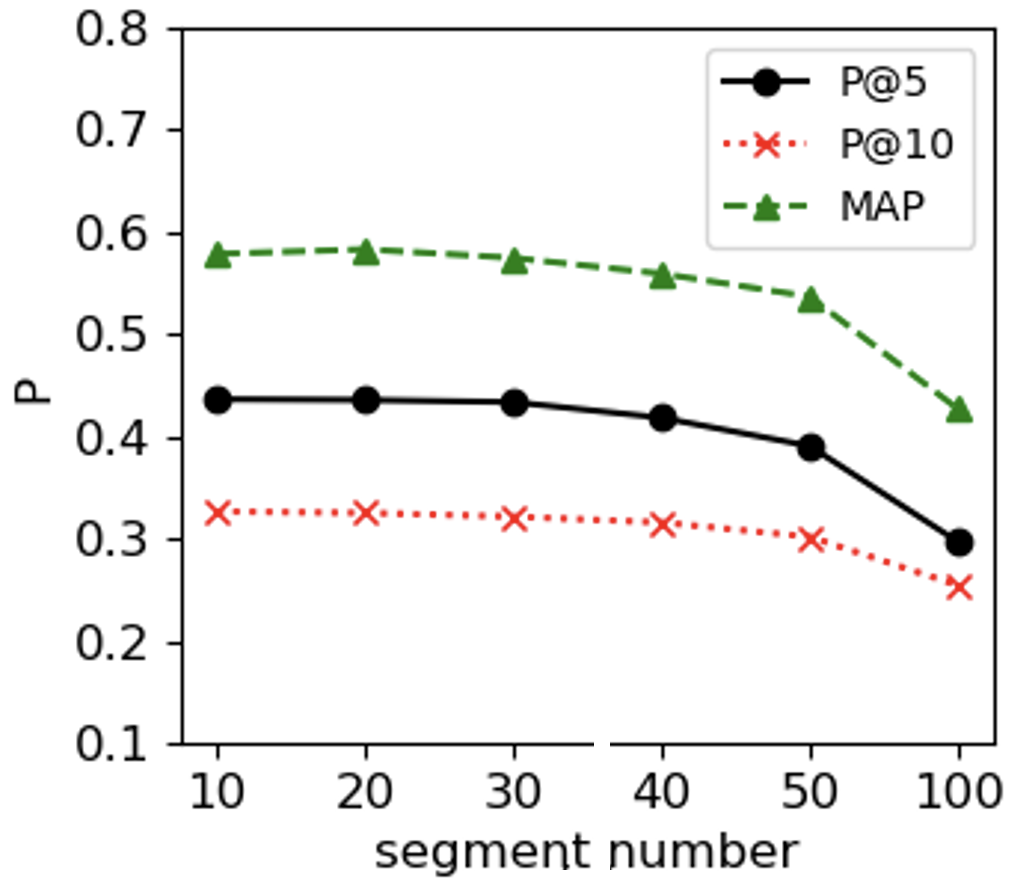}
    }
    \hfill
    \subfigure[]{
    \includegraphics[width=0.47\linewidth]{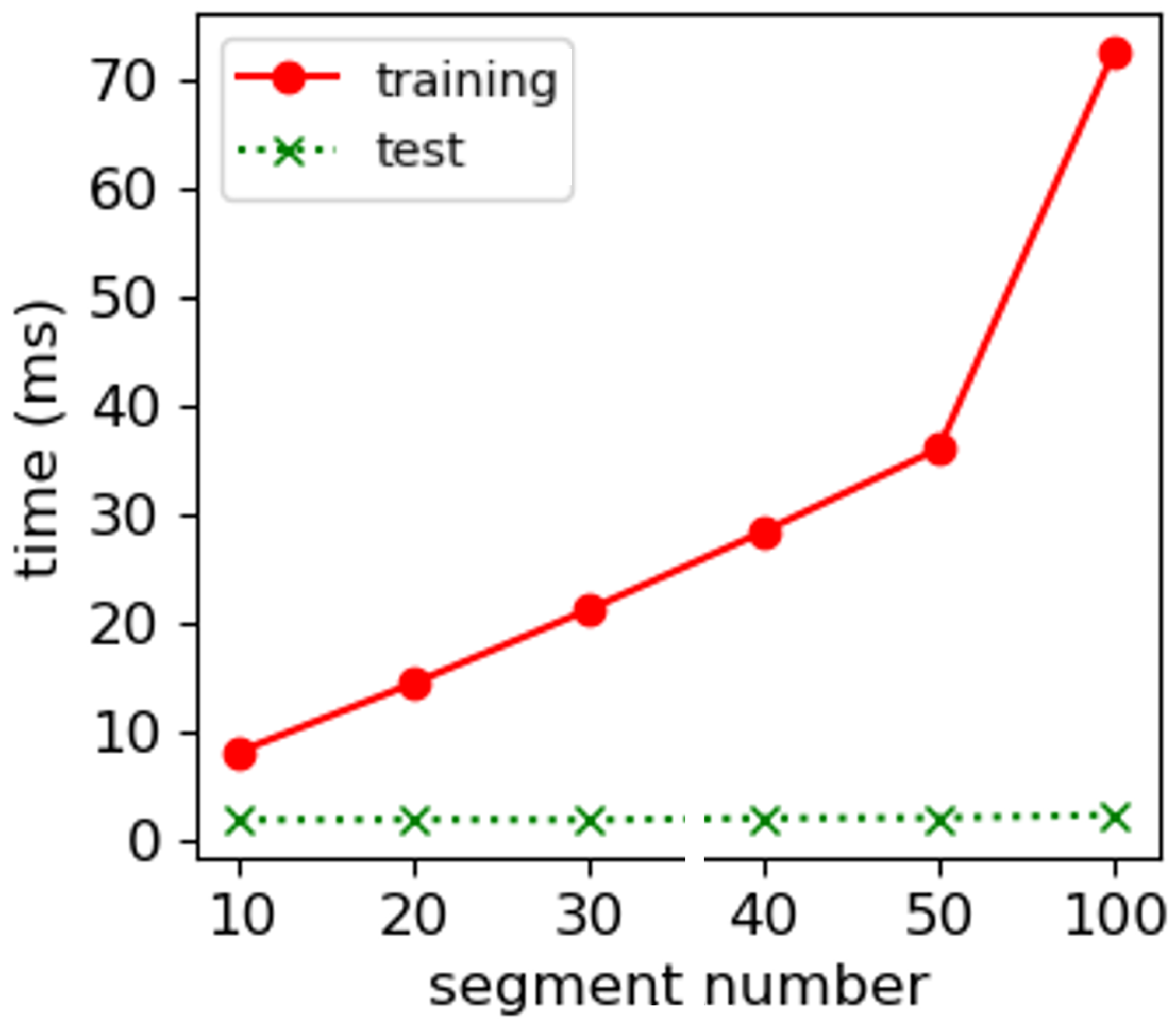}
    }  
    \caption{Effectiveness and Efficiency Results with different number of segments. Experimented with DeepTileBars+SEINE on MQ2008.} 
    \label{fig:kernel_size} 
\end{figure} 

\subsection{Main Results}


Table \ref{tab:result} reports the effectiveness and efficiency results for a few retrieval methods, such as KNRM, HiNT, and DeepTileBars, combined with different indexing methods on LETOR 4.0 first-stage document retrieval task. Our experiments demonstrated similar results for both MQ2007 and MQ2008. 
Here No Index is the baseline indexing method, which shows the performances when a retrieval method processes the entire document collection at query time without using any index. InvIdx  can only support bag-of-words retrievers and we show BM25 to illustrate.  SNRM and SEINE can both be used to support various retrieval methods. Note the original DeepCT system is included as one run under SEINE since we include DeepCT weights in our index.

Both SNRM and SEINE are able to significantly improve retrieval efficiency. SNRM makes the index sparse enough to handle a large corpus by mapping document text to new latent terms. Its sparsity helps reduce the amount of the interaction matrix calculation, thereby speeding up the retrieval phase. For instance, it gets 1.2$\times$, 1.2$\times$, 1.2$\times$ faster for training the KNRM, HiNT, DeepTileBars models, and 1.3$\times$, 1.1$\times$, 1.3$\times$ for test on the MQ2007 dataset. 
However, SNRM has a large degradation, ranging from -40\% to -9\% on the effectiveness metrics, over No Index. This might be caused by SNRM’s ineffectiveness in lexical matching, i.e., exact matching. SNRM is good at semantic matching, which aims to address a variety of linguistic phenomena, such as synonymy, paraphrase, and term variation; its index only stores the latent terms.  
Although it can be applicable to multiple neural retrievers as SEINE does, it is difficult for SNRM to get overlapping terms between query and document. Lexical information is lost in the latent semantic nodes. On the LETOR datsset, our results indicate SNRM makes a poor trade-off between efficiency and information loss.

On the contrary, SEINE stores the term-segment interaction in its index and generates a q-d interaction matrix by looking up and stacking rows for actual query terms. With minimal degradation in effectiveness, 
SEINE gets 3.7$\times$, 1.4$\times$, 7.4$\times$ faster for training the KNRM, HiNT, and DeepTileBars models, and 13.7$\times$, 1.4$\times$, 28.1$\times$ for test on the MQ2007 dataset. 
Similar results on the MQ2008 dataset can be observed.

\subsection {Impact of Segment Size}

We also look into the effects of different segment size in SEINE. We experiment with different number of segments per document using DeepTileBars+SEINE and report the findings on the MQ2008 dataset. 
Figure \ref{fig:kernel_size}(a) is about effectiveness. It shows that using 20 segments per document performs the best for precision (in fact for all other effectiveness metrics too).
Figure \ref{fig:kernel_size}(b) is about efficiency. It shows that as the number of segments per document increases, the neural networks take longer time to train while the test time remains more or less the same. We also find that 
the average segment length for the best choices of segments per document, 20 and 30, are around 270 and 200 words respectively. This length is about the length of a natural paragraph. It suggests that we should select our segment size close to an author's topical breaks, instead of choosing too large a segment size just for the sake of increasing training efficiency. At the same time, since test time does not vary much with segment size, we can select a smaller number of segments to improve query run-time efficiency.



\section{Conclusion}



This paper proposes SEINE, a SEgment-based Indexing method for NEural information retrieval, that moves heavy computations in interaction-based neural retrievers offline and only keeps essential calculations at query time. It provides a general indexing framework and flexibly supports a variety of interaction-based neural retrieval methods. During the indexing phase, we build a vocabulary from the entire corpus and compute and store the vocabulary-segment interactions in the index. We propose to use segment-level inverted index to store the atomic query-document interaction values. Our indexing method can make the query run-time up to 28 times faster without sacrificing their effectiveness on LETOR 4.0 for first-stage, full-length document retrieval. 


We propose to store atomic interactions between vocabulary term and segments in an inverted index. These interactions include term frequencies, inverse document frequency, dot products, operations over BERT embeddings, conditional probabilities, etc. Some of them are adopted from a few recent first-stage retrievers, such as DeepImpact~\cite{deepimpact}, EPIC~\cite{epic}, and TILDE~\cite{tilde}. 
Our experiments did not include them because (1) our main focus is to rejuvenate the index-less re-rankers that achieve high performance, but suffer in terms of efficiency.
However, (2) we do identify the atomic interaction function in DeepImpact~\cite{deepimpact}, EPIC~\cite{epic}, and TILDE~\cite{tilde}, and include them in our index. 
We hope re-building an index for neural retrieval can be avoided as much as possible so that  researchers can shift their attention back to  creating versatile retrieval functions. 

Currently, SEINE does not support MonoBERT, an all-in-all, dense interaction-based method, the most effective retriever at the moment. In all-in-all interaction, terms interact within and across a pair of query and document. To implement SEINE for such methods, we need to understand how to decompose these all-in-all interactions within the transformer blocks 
into some function of independent vocabulary entries. We leave this as future work.

\balance






\bibliographystyle{ACM-Reference-Format}
\bibliography{reference}


\end{sloppypar}

\end{document}